\documentstyle[emulateapj,pstricks]{article}

\def\mathnew{\mathsurround=0pt}
\def\ref{\par\noindent\hangindent=2pc \hangafter=1 }

\def\simov#1#2{\lower .5pt\vbox{\baselineskip0pt

\lineskip-.5pt\ialign{$\mathnew#1\hfil##\hfil$\crcr#2\crcr\sim\crcr}}}

\def\msun{\mbox{M$_{\odot}$}}

\def\Liso{\mbox{L$_{\rm iso}$}}

\def\flux{\mbox{photons cm$^{-2}$ s$^{-1}$}}
\def\pmin{\mbox{P$_{\rm min}$}}
\def\fldl{\mbox{$f$(L$_{\rm iso}$)dL$_{\rm iso}$}}
\def\sfrunit{\mbox{M$_{\sun}$Mpc$^{-3}$yr$^{-1}$}}
\def\ergs{\mbox{erg \ s$^{-1}$}}

\begin{document}

\slugcomment{{\em submitted to the Astrophysical Journal}}

\lefthead{Luminosity function and formation rate of GRBs}
\righthead{FIRMANI ET AL.}

\title{Formation rate, evolving luminosity function, jet structure, and 
progenitors for long Gamma-Ray Bursts}

\author{Claudio Firmani\altaffilmark{1}}
\affil{Osservatorio Astronomico di Brera, via E.Bianchi 46, I-23807 
Merate, Italy}
\affil{Instituto de Astronom\'{\i}a, U.N.A.M., A.P. 70-264, 04510, 
M\'exico, D.F., M\'exico}

\author{Vladimir Avila-Reese\altaffilmark{1}}
\affil{Instituto de Astronom\'{\i}a, U.N.A.M., A.P. 70-264, 04510,
M\'exico, D.F., M\'exico}

\author{Gabriele Ghisellini\altaffilmark{1}}
\affil{Osservatorio Astronomico di Brera, via E.Bianchi 46, I-23807 
Merate, Italy}
\and
\author{Alexander V. Tutukov\altaffilmark{1}}
\affil{Insitute of Astronomy, Russian Academy of Sciences,
Piatnitskaya st. 48, Moscow 109017, Russia}

\altaffiltext{1}{E-mail: firmani@merate.mi.astro.it, avila@astroscu.unam.mx,
gabriele@merate.mi.astro.it, atutukov@inasan.ru}


\begin{abstract}

We constrain the isotropic luminosity function (LF) and formation rate of 
long $\gamma$-ray bursts (GRBs) by fitting models {\it jointly} to both 
the observed differential peak flux and redshift distributions. We find 
evidence supporting an evolving LF, where the luminosity scales as 
(1+z)$^{\delta}$ with an optimal $\delta$ of 1.0$\pm$0.2. For a single 
power-law LF, the best slope is $\sim$ -1.57 with an upper luminosity 
of 10$^{53.3}$ \ergs, while the best slopes for a double power-law LF 
are approximately $-1.6$ and $-2.6$ with a break luminosity of 10$^{52.7}$ \ergs.
Our finding implies a jet model intermediate between the universal structured 
$\epsilon(\theta)\propto \theta^{-2}$ model and the quasi-universal Gaussian 
structured model. For the uniform jet model our result is compatible with 
an angle distribution between 2$^\circ$ and 15$^\circ$.
Our best constrained GRB formation rate histories increase 
from z=0 to z=2 by a factor of $\sim$ 30 and then continue increasing
slightly. We connect these histories to that of the cosmic star formation 
history, and compare with observational inferences up to z$\sim$6. GRBs could 
be tracing the cosmic rates of both the normal and obscured star formation 
regimes.  We estimate a current GRB event rate in the Milky Way of
$\sim 5 \ 10^{-5}$ yr$^{-1}$, and compare it with the birthrate of massive 
close WR+BH binaries with orbital periods of hours. The agreement
is rather good suggesting that these systems could be the progenitors
of the long GRBs.
\end{abstract}

\section{Introduction}

The finding of the luminosity function (LF) and formation rate history (FRH) of 
long $\gamma$-ray bursts (GRBs) is a key issue to understand the nature of these 
events, as well as to use them as potential tracers of the star formation rate 
(SFR) in the universe. The most direct way to infer both the LF 
and FRH is based on the observational luminosity-redshift diagram (LZD) 
(Schaefer, Deng \& Band 2001; Lloyd-Ronning, Fryer \& Ramirez-Ruiz 2002; 
Yonetoku et al. 2003). However, this strategy is conditioned by (i) the bias 
introduced on the sample by the redshift inference method, and (ii) the 
sensitivity limit related to the redshift estimate, which in current LZDs 
is $>1$ \flux. In order to find a reasonable LF, this limit has to be pushed 
down at least by an order of magnitude. These goals will be reached by future 
space missions; for example, the {\it Burst Alert Telescope, BAT} on-board 
{\it Swift}\footnote{http://swift.gsfc.nasa.gov/science/instruments/} 
will be 5 times more sensitive than BATSE, reaching a limiting flux of 
10$^{-8}$ erg cm$^{-2}$ s$^{-1}$.  Another approach, 
commonly used in the past, is based on fitting models with parametric LFs 
and with a given FRH to the observed peak flux GRB distribution, 
for which the statistics is healthy (see Stern et al. 2002a, and  Guetta, 
Piran \& Waxman 2003, and more references therein). Nevertheless, 
the complicated mixing between the model LF and FRH introduces a degeneracy 
among these factors in the flux distribution (e.g., Krumhol, Thorsett \& 
Harrison  1998).
Some early attempts to fix extra constraints on the fits by using
the information provided by the few GRBs with known redshifts were done by 
Sethi \& Bhargavi (2001), Stern et al. (2002a,b) and Guetta et al. (2003).

Taking into account the difficulties mentioned above, here we propose a new 
strategy to constrain the GRB LF and FRH from the available observations. 
We will use the widest sample up to date for the {\it differential} peak-flux, P, 
distribution (the logN-logP diagram, NPD hereafter) (Stern et al. 2002b: SAH02), 
as well as the differential redshift distribution (the N-z diagram, NZD 
hereafter) from 220 GRBs inferred from an empirical luminosity-indicator 
relationship (Fenimore \& Ramirez-Ruiz 2000: FR00) or from the sample of 33 
GRBs with known redshifts (e.g., see the compilation given in van Putten \& 
Riegenbau 2003: PR03). Then, through Monte-Carlo simulations we will 
optimize the best LF and FRH models by fitting both distributions 
(NPD+NZD) {\it simultaneously} by an accurate minimum global $\chi^2$ 
criterion.

After the discovery of afterglows for the long GRBs, it was realized that their 
emitted luminosity should be collimated. The simplest model for the jet implies 
a {\it uniform} energy distribution across the jet and a sharp drop outside 
$\theta_j$, where $\theta_j$ is the opening angle related to the observed 
achromatic break time in the light curves (Rhoads 1997). 
For those bursts with known redshift, and therefore with known {\it isotropic
luminosity} \Liso, \Liso$\theta_j^2$ is nearly constant  (Guetta et al.
2003), although its distribution results wider than the one found
for E$_{\rm iso}\theta_j^2$ (Frail et al. 2001; Bloom, Frail \& Kulkarni 2003).
A further {\it structured} jet model proposes a universal non-uniform energy 
distribution within the jet, the observed diversity of break times in the light 
curves being caused by a variation of the viewing angle $\theta_v$ 
(Postnov, Prokhorov \& Lipunov 2001; Rossi, Lazzati \& Rees 2002; Zhang \& 
M\'esz\'aros 2002). 
Similar to the uniform jet model, at least for an energy distribution per 
solid angle $\epsilon(\theta_v)\propto \theta_v^{-2}$, it was found that 
\Liso$\theta_v^2$ is also roughly constant. An alternative approach 
supposes a Gaussian energy distribution with angle (Zhang  \& 
M\'esz\'aros 2002; Kumar \& Granot 2003).
Thus, {\it the origin of the isotropic LF of GRBs could be due in large 
part to the diversity of opening or viewing angles}. 
Therefore, any improvement on the LF knowledge means a progress on the
understanding and constraining of the jet structure (e.g., Guetta et al.
2003; Nakar, Granot \& Guetta 2003).   

The {\it collimation} of the luminosity is also important at the time 
to calculate the enhancement factor between the {\it observed} and the 
{\it true} event rates of GRBs. The enhancement factor for the uniform 
jet model has been estimated previously to be $<$500 (Frail et al. 2001; 
PR03), while for the structured jet model, values smaller by more than
an order of magnitude were obtained (Zhang et al. 2003; Guetta et al. 2003). 
Guetta et al. (2003) have also found a small value for this factor
($75\pm25$) in the case of the uniform jet model. Unlike 
Frail et al., Guetta et al. used a weighted average of the 
angular distribution.

An ultimate goal of astronomical studies on GRBs is the connection of the
GRB FRH to the SFRH in the universe. Several pieces of 
evidence show that long GRBs are associated to the core collapse of massive
stars. Since the detected $\gamma$-ray fluxes may 
come from very high-redshifts and from dust enshrouded media, GRBs may offer 
an interesting way to explore massive star formation in galaxies without 
the uncertainties of dust extinction, as well as in the universe up to very high 
redshifts (e.g., Totani 1999; Wijers et al. 1998; Lamb \& Reichart 2000; 
Blain \& Natarajan 2000; Ramirez-Ruiz, Threntan \& Blain 2002; 
Choudhury \& Srianand 2002; Bromm \& Loeb 2002). 

Here we will attempt to constrain the GRB FRH by using both the NPD and NZD, 
and devoting special attention to the behavior at z$>$2. We will also make
use of our results to explore the nature of the GRB progenitors. The association 
of long GRBs to SNIb/c narrows the diversity
of potential progenitors to massive helium (Wolf-Rayet, WR) stars,
which, owing to the angular momentum requirement, should be in a 
close binary system with a compact companion. An important task is to
estimate the birthrate of these systems in the Milky Way and compare it 
with the birthrate inferred for the GRBs. 

In \S 2 we describe the observational data that will be used in our study.
In \S 3 the model and strategy applied to constrain the LF and FRH from 
observations are described.  
In \S 4, results regarding the best fitting LFs and FRHs are presented;
evolving LFs are strongly favored, and the FRH shape is found to 
be topologically similar to the global SFR history (SFRH). 
In \S 5 we discuss the implications of an evolving LF on current jet models, 
while in \S 6 the connection between GRB FR and SFR, and some implications 
for the progenitors are discussed. 
Finally, our conclusions are given in \S 7.
Throughout this paper we use the flat $\Lambda$CDM cosmology with 
$\Omega_m$=0.29, $\Omega_{\Lambda}$=0.71, h=0.71.

\section{The Data}

Stern et al. (2001,2002a) have processed a sample of 3255 BATSE triggered 
and non-triggered GRBs longer than 1 s. By using  an appropriate efficiency 
matrix they were able to extend the peak flux limit down to 0.1 \flux. In 
order to extend the flux distribution in the bright side, the data from 
{\it Ulysses} satellite have been used by SAH02 by doing a cross-calibration 
of the joint {\it Ulysses/BATSE} events. The final result is the 
{\it differential} peak-flux distribution corrected by a joint {\it Ulysses/BATSE}
exposure factor that extends from P=0.1 to about 300 \flux. 
This is the NPD that we use here (kindly made available in electronic form 
to us by B.E. Stern). In some previous works, the {\it cumulative} P distribution
has been used. As noticed by the referee, random errors in a cumulative 
distribution propagate in an unknown way, therefore for the analysis we 
pretend to do, the differential distribution should be used. 

For the differential redshift distribution we use two samples. The first one is a 
set of 220 BATSE GRBs with a sensitivity threshold of 1.5 \flux\ and with redshifts 
inferred by using the luminosity-variability empirical relation derived by FR00 
(kindly made available in electronic form to us by E. Ramirez-Ruiz). 
The second sample is a set of 33 GRBs with known z 
taken from PR03 (excluding GRB980425 and adding GRB030429). This sample is 
highly biased by selection effects. We attempt to correct some of these effects 
following Bloom (2003). We use the probability of redshift detection due to 
observability of lines in the spectral range and in the presence of night-sky 
lines estimated by Bloom (2003) (who is kindly acknowledged by making available 
for us an electronic table with the probabilities).
The effects related to the distance are accounted with another probability which 
we set equal to 1 for z$\le$ 1 and decreasing as 1/(1+z) for larger
redshifts (see Gou et al. 2003 for the flux detection of GRB afterglows 
located at different redshifts).

\section{The Model}

The differential NPD is modeled by {\it seeding} a large number of GRBs with 
a given FR, $\dot{\rho}_{\rm GRB}$ (per unit of comoving volume), and  LF, 
\fldl, and by propagating the flux of each source to 
the present epoch. The rate of GRBs observed with peak fluxes between 
P$_1$ and P$_2$ is ($\propto$ NPD):
\begin{equation}
{\rm \dot{N}(P_1,P_2)=\int_0^{\infty}dz\frac{dV(z)}{dz}
\frac{\dot{\rho}_{GRB}(z)}{1+z} \int_{L_1}^{L_2} \fldl}, 
\end{equation}
with L$_1$=\Liso(P$_1$,z) and L$_2$=\Liso(P$_2$,z), while  
the rate of GRBs observed between z$_1$ and z$_2$ ($\propto$ NZD) with a 
limiting sensitivity \pmin\ is:
\begin{equation}
{\rm \dot{N}({z_1},{z_2})=\int_{z_1}^{z_2}dz\frac{dV(z)}{dz}
\frac{\dot{\rho}_{GRB}(z)}{1+z} \int_{L_m}^{\infty}\fldl}, 
\end{equation}
where L$\rm_m$=\Liso(\pmin,z), dV(z)/dz is the comoving volume at z, 
the term 1/(1+z) 
takes care of time dilation, and \Liso(P,z) is obtained by solving the equation:
\begin{equation}
{\rm P=\frac{(1+z)\int_{(1+z)E_{min}}^{(1+z)E_{max}}S(E)dE}{4{\pi}d_L^2(z)}},
\end{equation} 
where E$_{\rm min}$=50 keV and E$_{\rm max}$=300 keV define the sensitivity 
band, S(E) is the source differential rest-frame photon luminosity, 
and d$\rm_L$(z) is the luminosity distance (see Porciani \& Madau 2000). 
\Liso\ in the rest frame is defined as 
\Liso=$\int_{\rm 30keV}^{\rm 10000keV}$ES(E)dE. 
For the spectrum, we use the Band et al. (1993) proposed fit to a sample 
of BATSE spectra, with the parameters taken from averages re-derived by 
Preece et al. (2000) ($\alpha=-1$, $\beta=-2.25$, E$\rm_b=256$ keV). 
Notice that these values are in the observer rest frame. 
It is assumed that the shape of the Band et al. spectrum does 
not change with z. In the assumption that the average redshift of the GRB 
sample used here is z=1, E$\rm_b$ would be 511 keV (Porciani \& Madau 2000). 

Recently it was discovered a relation between the observed peak energy in the 
spectrum (related to E$\rm_b$) and \Liso\ (Yonetoku et al. 2003; see 
Amati et al. 2002 and Lamb, Donaghy \& Graziani 2003 for the analogous 
correlation between E$\rm_b$ and E$_{\rm iso}$). We will explore both cases, 
a rest E$\rm_b$ constant and a rest
\begin{equation} \label{Amati}
{\rm E_b \simeq 15\ (\Liso/10^{50}\ \ergs)^{0.5}\ keV}.           
\end{equation}

Once defined the cosmological model, the LF, and the FRH, then we calculate the 
NPD and NZD trough Monte-Carlo simulations. We use a single power law (SPL) LF,
\begin{equation} \label{lumfun1}
f(\Liso)\propto \Liso^{-\gamma},\ \ \ \ \ \ \ {\rm L_l<\Liso<L_u},          
\end{equation}
supported by several theoretical models, or a double power law (DPL) LF:
\begin{equation} \label{lumfun2}
f(\Liso)\propto \left\{ \begin{array}{ll}
              \Liso^{-\gamma_1},   &  \mbox{$\ \ \rm L_l<\Liso<L_b$} \\\\
              \Liso^{-\gamma_2},   &  \mbox{$\ \ \rm L_b<\Liso<L_u$}
           \end{array}
   \right.           
\end{equation}
suggested by several studies (e.g., Schmidt 1999,2001; FR00; Schaefer et al. 
2001; SAH02). From a statistical analysis of a sample of GRBs with 
z and \Liso\ inferred from empirical luminosity-estimator relationships,  
Lloyd-Ronning et al. (2002) and Yonetoku et al. (2003), have suggested that 
the luminosity should scale with redshift as (1+z)$^{\delta}$, 
with $\delta \approx$1.4 and 1.9, respectively.

The GRB FRH per unit of comoving volume is given by: 
\begin{equation} \label{formrate1}
{\rm \dot{\rho}_{GRB}= K \dot{\rho}_{SF}}, 
\end{equation}
where $\dot{\rho}_{\rm SF}$ is the GRB FRH normalized to a level close to 
the one of cosmic SFRH, and defined as:
\begin{equation} \label{formrate2}
{\dot{\rho}_{\rm SF}= {\eta}({\rm z})\ \frac{0.3\ e^{b{\rm z}}}{e^{b{\rm z}}+a}
 \ \sfrunit}. 
\end{equation}
From observations of cosmic SFRH, Porciani \& Madau (2001) fit $a=22$ and $b=3.4$. 
Varying $a$, this equation allows to study the effects of the SFR decline 
from z=2 to 0. The function $\eta$(z)=1 for z$<$2, and $\eta$(z)=e$^{c({\rm z}-2)}$ 
for z$>$2, allows to study the effects of a growth (decline) of 
$\dot{\rho}_{\rm SF}$ for  z$>$2 depending on the value of $c>0$ ($c<0$).   
Notice that the factor K gives information on the GRB
progenitor FR and has unities of $\msun^{-1}$.  

The strategy is to constrain the LF parameters ($\gamma$,L$\rm_u$) for SPL (keeping 
L$\rm_l$=10$^{49}$ \ergs), and ($\gamma_1$, L$\rm_b$, $\gamma_2$) for DPL (keeping 
L$\rm_l$=10$^{49}$ \ergs\ and L$\rm_u$=10$^{55}$ \ergs), as well as the $a,\ b,\ c,
\ \delta$, and K parameters. 
The constraints will be obtained by a joint fit of the model predictions to the 
observed NPD and NZD (including their errors), using a global 
$\chi^2$ criterion. The total $\chi^2$ is the sum of $\chi_{\rm NP}^2$ and  
$\chi_{\rm NZ}^2$. In order to verify the quality of fit, tests have been 
made assuming the total $\chi^2$ to be a linear combination of $\chi_{\rm NP}^2$ 
and $\chi_{\rm NZ}^2$, with weight coefficients that scale inversely to the 
number of data (bins) in the NPD and NZD samples, respectively, in such a way 
that each sample conditions the model with the same strength. The parameter 
optimization is based on the non linear Levemberg-Marquardt method to find 
the least $\chi^2$ (Press et al. 1988). In spite of the large uncertainties, 
the redshift distribution of 33 GRBs with known z will be also used as a 
complementary test.

\section{Results}

We have run several models with a SPL or DPL LF (identified by S an D, 
respectively), adopting E$\rm_b$=511 keV or E$\rm_b$ as a function 
of \Liso\ according to eq. (4) (Yonetoku et al. 2003; 
identified by P and Q, respectively), and either without evolution ($\delta=$0) 
or including $\delta$ among the parameters to optimize (identified by 0 and E, 
respectively). The parameters of each one of these models were optimized
by a simultaneous fit to the observed NPD and NZD samples, as described 
in \S 3. Table 1 gives for each model, identified as was 
specified above, the values of the fitted parameters, where the same symbols 
of \S  3 were used. The uncertainties are the conventional standard deviations.
The number of degrees of freedom (for the simultaneous fit) corresponding
to each models given in Table 1 is 43 minus the number of parameters 
shown in Table 1 for the corresponding model.  The information about the 
quality of the fit for the models is given in Table 2. For each model, this 
Table shows the total $\chi^2$, the goodness-of-fit test given by the 
probability Q to find a new total $\chi^2$ exceeding the current value, 
the partial $\chi_{\rm NP}^2$ and $\chi_{\rm NZ}^2$ as well as the significance 
levels, $\rm P_{\rm NP}^{\rm KS}$ and $\rm P_{\rm NZ}^{\rm KS}$, of the 
Kolmogorov-Smirnov test between the fitted distributions and the NPD and NZD 
samples, respectively. We have introduced the Kolmogorov-Smirnov test just for 
completeness; in fact, for all the models the obtained values of the significance 
levels are not sufficiently high as to disprove the null hypothesis that the 
expected and observed cumulative distributions are from the same parent 
distribution. Nevertheless, even if these significance levels are affected by 
noise, they reveal a minor quality of the fit for SP0 and DP0 models as 
well as an intermediate quality for SQ0 and DQ0 models. The Q probability
shows clearly this situation.

In Fig. 1 we plot the observed and the modeled NPD and NZD. The models with SPL 
and DPL LFs are on top and bottom panels, respectively. Dotted lines are for models 
without evolution (0), while solid lines are for models with the optimized 
evolution parameter $\delta$ (E). Thin and thick lines identify the cases P 
and Q, respectively. The data are the points with error bars from SAH02 (NPD) 
and  from FR00 (NZD). \\

{\bf GRB luminosity function and evolution.}
The fit of models to the data at the side of the lowest P's in the NPD fixes 
the value of the K factor, while the slope here determines the $\gamma$ (SPL) 
or $\gamma_1$ (DPL) LF parameters. The quality of the fit can be estimated 
actually at intermediate and high P's. A scale invariant (without cutoff) 
power law LF should provide a (log-log) NPD close to a straight line
because the power index keeps memory in the NPD, while the FRH does not
influence on it (e.g., Loredo \& Wasserman 1998). Any break to scale invariance 
introduces some curvature besides of the volume evolution effect. 
The imprint of the FRH on the NPD becomes more and more significant as the LF 
diverges from scale invariance. On the other hand, if the LF evolves, 
then as $\delta$ increases, the LF tends to concentrate again its influence 
on the NPD, and the curvature in the NPD grows.
This is evident in the NPDs of Fig. 1. Non-evolving LFs show a 
low NPD curvature because L$\rm_u$ or L$\rm_b$ for GRBs at different z spread 
on a large P range, while for increasing evolutionary effects this spread is 
reduced and the curvature grows. Finally, it is quite obvious that a 
reduced range in P (e.g., 1-50 \flux) weakens strongly the condition given by 
the NPD, and no robust predictions can be made; the range used here is rather 
large, 0.1-300 \flux. The results regarding non-evolving and evolving LFs are:

1. For both the SPL and DPL cases, the {\it non-evolving LFs} fit poorly 
the data  as one may appreciate by the values of the $\chi^2$, Q and 
$\rm P^{\rm KS}$ of Table 2. 
In the NPD, the non-evolving models show an excess of GRBs at high P, 
while in the NZD their fits to the data appear by far to be the best ones
(Fig. 1). In terms of our comparative analysis, 
{\it the non-evolving LF can be ruled out}. 

2. The quality of the fit becomes excellent when an {\it evolving LF} 
is adopted. The evolutionary models indicate an optimal value of 
$\delta={\rm 1.0} \pm {\rm 0.2}$. Care has to be taken, however, on the 
fact that the LF is sensitive also to the constraints on the NZD.
Therefore, a careful analysis has to be done, using NZDs obtained from samples 
based on different z determinations, in order to achieve a more conclusive 
value for $\delta$. 

3. The application of our method to the observational NPD and NPZ samples
used here points out to a marginal preference of the SPL LF on the DPL LF. 
For a more conclusive result on this question, an analysis on a complete 
LZD has to be carried out.

Taking into account these considerations, our preliminary conclusion is that {\it 
an evolving LF seems to be necessary in order to reproduce both the NPD and NZD.} \\ 

{\bf GRB formation history.}
The factor K connects the GRB FR (frequency) to the {\it assumed} SFR, 
and it is fixed mainly by the side of low P's in the NPD. Notice that in 
Eq. (8), $\dot{\rho}_{\rm SF}$ was normalized to about 0.2 \sfrunit\ at z=2 
for most of the models (see Fig. 2); a different normalization would imply 
an equivalent inverse scaling of K. It is important to remark that the GRB 
FRH takes into account the contribution of the overall events above 
L$\rm_l$ at each z (according to the given LF), while in the NZD the lower 
limit is determined by the sensitivity threshold on \pmin. As seen in 
Fig. 1, the best models in the NZD are those with an evolving LF. 
Figure 2 shows the SFRHs corresponding to the models of Fig. 1, using the 
same line code. The vertical segmented area shows the statistical uncertainty 
of the Q evolving models. The result is encouraging:  
{\it the GRB FRHs best constrained from observations follow, at least 
qualitatively, the observed cosmic SFRH up to z$\approx$6 (see \S 6 and Fig. 4)}, 
showing a steep increase from z=0 to z$\sim$2 (a factor of $\sim$30) and a 
a gentle increasing towards earlier epochs. This result shows 
the potential usefulness of GRBs as tracers of SFRs at high redshifts. 
In fact our results suggest some difference among the SFRHs
inferred from GRBs and rest-frame UV luminosity (Fig. 4). This difference 
is expected if the GRB-based SFRH is tracing both normal and obscured SF regimes
(see \S 6).  \\

The effect on the results of assuming whether a constant E$\rm_b$ (case P) or 
an E$\rm_b$ depending on \Liso\ according to Eq. (4) (Yonetoku et al. 2003; 
case Q) is not relevant. For models with a non-evolving LF the fits
improve a little when using E$\rm_b$ depending on \Liso, but their quality
remain in any case much worse than that of the models with an evolving LF.
In our calculations we have assumed, rather arbitrarily, the low luminosity 
cut-off L$\rm_l$=10$^{49}$ \ergs. Actually the fit is invariant with 
respect to a decrease of L$\rm_l$ below $\sim 10^{50}$ \ergs, mainly because the 
GRBs with lower luminosities appear above the sensitivity limit of 
0.1 \flux\ in a very reduced volume around the observer. The only fitting 
parameter that changes with L$\rm_l$ is K, which scales as L$_{\rm l}^{-0.6}$ 
for the best models. The highest correlation between the parameters appears 
between Loga and b, for which the Pearson's correlation coefficient is 0.85 
and $\Delta$Loga $\approx$ 0.4 $\Delta$b.

Finally, we have also carried out our analysis using the NZD from the de-biased
sample of 33 GRBs with known redshifts (see \S 2). Some difficulty derives from the
fact that the sensitivity threshold, \pmin, of this sample is unknown. 
Therefore, we have experimented choosing different thresholds 
between 1 and 5 \flux, and proved that the changes on LF and FRH 
are not significant. The constrained SFRH and its
statistical dispersion are shown in Fig. 2 by the 
shaded area, assuming {\pmin}=5 \flux, and $\delta=1$. A 
qualitative agreement with results obtained using the NZD from FR00 is seen. 
Thus, {\it from different observational samples we reach the same
conclusion: the shape of the GRB FRH approximates that one of the cosmic 
SFRH}.

\section{Luminosity function and jet models}

As mentioned in the introduction, several pieces of evidence suggest that 
the LF of GRBs could be strongly related to jet collimation effects. From our
analysis, using the observed NPD and NZD, we have obtained $\gamma=1.55\pm 0.05$
for the SPL model. This result provides some evidence against the
universal structured jet model with $\epsilon(\theta)\propto \theta^{-2}$ (see
also Guetta et al. 2003; Nakar et al. 2003), which predicts a differential LF 
$\propto$L$^{-2}$ (Rossi et al. 2002; Zhang \& M\'esz\'aros 2002), and also against 
the quasi-universal Gaussian jet structured model, which predicts a differential 
LF $\propto$L$^{-1}$ even if some dispersion of parameter values is allowed
(Lloyd--Ronning, Dai \& Zhang 2003). 
Notice, however, that the best slope we have obtained is intermediate 
between these two cases. In the framework of structured jet models, the low 
luminosity branch of the LF is testing how the power per unit solid angle 
behaves for relatively large angles (i.e. corresponding to small apparent 
luminosities). For power law structured jets, with 
$\epsilon(\theta)\propto \theta^{-k}$, the predicted LF is a power law 
with slope $\gamma$=1+2/k (Zhang \& M\'esz\'aros 2002). Therefore, 
$\gamma$=1.5 corresponds to k=4, i.e. a steeper value than the ``canonical" 
k=2. This behavior might correspond to the wings of the jet, and not to the 
entire jet structure: after all, to avoid divergence, $\epsilon( \theta )$ 
needs to be a much more moderate function of $\theta$ for small $\theta$. 
Note also that a power law description of $\epsilon(\theta)$ may be an 
oversimplification of the real case, which could be described instead by 
a double Gaussian profile (one for the core of the jet and another for the 
wing, as suggested by the behavior of the GRB 030329 light-curve, Berger 
et al. 2003a). Also, the unification of GRB and XRF requires $\epsilon(\theta)$ 
to fall off more steeply than $\theta^{-2}$ at large angles, to not 
over-produce XRFs (Lamb et al. 2003; Zhang et al. 2003).

For a uniform jet, \Liso=2L$_0$/$\theta_j^2$  ($\theta\leq\theta_j$) and 
\Liso=0   ($\theta > \theta_j$), being $\theta_j$ the beam aperture
and L$_0$ the intrinsic GRB luminosity (Guetta et al. 2003; see also
Frail et al. 2001; Bloom et al. 2003). 
The differential isotropic luminosity distribution scales with $\theta_j$ 
as dN$\propto\theta_j^2$ dN$_j$, where dN$_j$ is the differential jet 
angle distribution. Let assume dN$_j\propto\theta_j^{-\nu}$d$\theta_j$ and 
dN$\propto \Liso^{-\gamma}$ d$\Liso$. With some elementary algebra 
one obtains that $\nu=5-2\gamma$ (see also Guetta et al. 2003). 
Thus, in the case of the SPL LF the opening angle distribution is: 
dN$_j$=0  ($\theta_j<{\rm (2L_0/L{_u})^{1/2}}$) 
and dN$_j\propto\theta_j^{-1.9\pm0.1}$d$\theta_j$ 
($\theta_j \geq {\rm (2L_0/L_u)^{1/2}}$), 
while in the case of a DPL LF, the opening angle distribution is: 
dN$_j\propto\theta_j^{1.0\pm 1}$ d$\theta_j$  
($\theta_j<{\rm (2L_0/L_b)^{1/2}}$) and 
dN$_j=\theta_j^{-1.95\pm0.06}$d$\theta_j$ 
($\theta_j \geq {\rm(2L_0/L_b)^{1/2}}$).

Now, taking into account the bursts of known redshift, 
peak flux and jet angles (from PR03 and Bloom et al. 2003), 
at z$\sim$1 we estimate an average value of 
2L$_0\sim$0.8 10$^{50}$ erg/s (see Fig. 3) and 
$\theta_b \equiv\ (2L_0/L{\rm_b})^{1/2} \sim$ 2.3$^\circ$. 
The upper limit $\theta_u$ of the jet angle distribution 
should be determined by the lower limit of LF. 
This is not our case because of the NPD is not 
sufficiently extended toward low peak fluxes. 
Then we adopt an upper 
limit obtained by the \Liso=${\rm 2L_0}/\theta_j^2$ relation, the 
Eq. (4), and a low limit E$\rm_b \sim$ 50 keV on the ground of the BATSE
sensitivity. 
This assumption gives $\theta_u \sim$ 15$^\circ$. Based on 
geometrical considerations, values for $\theta_u$ up to 60$^\circ$ cannot 
be excluded. For the DPL LF case the enhancement factor 
f$_e = \int$dN$_j$/$\int$ dN gives f$_e$=3/$(\theta_b \theta_u) \sim$ 280, 
the uncertainty being at least a factor two.  

Our results suggest a mild luminosity evolution, L$\rm_b$$\propto$(1+z)$^\delta$ 
with $\delta \sim 1$. We have checked that this luminosity evolution 
does not contradict the suggestion of a universal energy reservoir for 
GBRs (Frail et al. 2001). To this end we have calculated the {\it observed} 
peak luminosities of the bursts listed in PR03 using a conversion factor 
between the listed count rate and flux of 1 count cm$^{-2}$ s$^{-1}$ = 8.7 
10$^{-8}$ erg cm$^{-2}$ s$^{-1}$. For the bursts in common with Bloom et al.
2003), we could then calculate the {\it true} peak 
luminosity using the jet opening angles listed in that paper. Figure 3 
shows the jet opening angles and the corresponding {\it true} energies, as a 
function of redshift, for the 24 bursts listed in Bloom et al. (2003).
This figure shows also the {\it true} luminosity for the 19 GRBs listed
both in PR03 and Bloom et al. (2003). 
For illustration, the dashed line shows the case of  
an evolution of the {\it true} luminosity proportional to (1+z). 
It is clear that this evolutionary behavior does not 
contradict the existing data. Note also that the found evolution of the 
{\it observed} luminosity could be associated not to the evolution of the 
{\it true} luminosity, but to the evolution of the aperture angle of the jet.

\section{GRBs as tracers of the cosmic Star Formation Rate and 
implications for the progenitors}

The understanding of the SF processes and history is a key issue to 
integrate a global vision of the universe.  
GRBs offer the hope of a deep insight of these 
processes if we will be able to establish a connection between GRBs 
and stellar evolution. Our results, though still uncertain, have shown 
that the GRB FRH resembles qualitatively the observed cosmic SFRH.
The SFRH traced by the rest-frame UV luminosity and corrected by dust 
obscuration, as presented in Giavalisco et al. (2004), is shown in Fig. 4.
In this figure we plot also our best models from Fig. 2. 
A significant contribution to the cosmic SFRH could come from sources 
emitting strongly in the rest-frame far infrared/submillimetre 
(e.g., Blain et al. 1999; Dunne, Eales \& Edmunds 2003). These objects 
are likely dust enshrouded SF bursts induced by galaxy collisions that 
follow the major mergers of dark halos at earlier epochs (for a
recent review on galaxy formation, see e.g., Firmani \& Avila-Reese 2003). 
GRBs could be direct tracers of the 
SFRH of these galaxies, although this is still an open question 
(see e.g., Ramirez-Ruiz et al.  2002; Berger et al. 2003b; Le Floc'h et
 al. 2003; Barnard et al. 2003). 

The possibility that GRBs are tracing also the SFRH in obscured galaxies 
might explain the apparent differences between the inferred GRB FRH and 
the observed UV-cosmic SFRH in Fig. 4. Nevertheless, our results are still 
uncertain due to the nature of the data used to construct the NZD, and are 
not suitable for a quantitative exploration on these aspects. In the 
future, when more data will be collected, we hope that the comparison 
of the SFRHs inferred from luminous SFR tracers and from the GRB FRs 
will help to clarify interesting questions related to the nature of 
the GRBs as well as to the SF processes in highly-obscured galaxies and 
in the high-redshift universe, where reionization makes difficult the 
observability of typical emission lines associated to SF.  

The value of K$\equiv \dot{\rho}_{\rm GRB}/ \dot{\rho}_{\rm SF}$ (see Table 1), 
gives the number of {\it observable} (beaming selected) GRB per unit of gas 
mass transformed into stars. For the preferred models, K$\approx$ 
0.5 10$^{-7}$ M$_{\sun}^{-1}$. This value is uncertain because it depends 
on the assumed $\dot{\rho}_{\rm SF}$ normalization and on the lower LF limit 
L$\rm_l$. 
For a present-day SFR of 4 \msun yr$^{-1}$ for the Milky Way, we obtain an event 
rate of $\sim$ 2 10$^{-7}$ yr$^{-1}$, which should be enhanced by the factor 
f$\rm_e$=280 (see \S 5); the {\it true} event rate is then $\sim$ 5 10$^{-5}$ 
yr$^{-1}$. This rate is rather uncertain, at least by a factor 2. It is about 
fifty times lower than the frequency of SNIb/c in the Milky Way, showing 
that only a small fraction of WR stars exploding as SNIb/c give rise to the GRB 
phenomenon (see also Podsiadlowski et al. 2004).
      
Within the framework of the popular collapsar model for long GRBs, 
the high angular momentum requirement is the main difficulty (e.g., 
Zhang \& M\'esz\'aros 2003). In fact, the strong mass loss by the star 
and the core-to-envelope (dynamo) magnetic coupling slow down 
the core rotation (Woosley, Zhang \& Heger 2002; Spruit 2002; Heger \& 
Woosley 2002; Izzard, Ramirez-Ruiz \& Tout 2003). 
Instead, if the pre-supernova is in a binary system, then the 
spin-orbit tidal interactions may increase the rotation even to the limit 
of Kerr black hole (BH) formation after SNIb/c explosion, ensuring a 
centrifugally supported disk.

Binary systems with a WR star co-rotating with the orbital motion and with 
a period of hours (which implies a massive BH as the companion) could be the 
progenitor of long GRBs (Tutukov \& Cherepashuk 2003; Tutukov 2003). In fact, 
the condition for Kerr BH formation is $\omega_K R_K \sim c$, where $\omega_K$, 
and $R_K$ are the angular velocity, and the radius of the formed Kerr BH, 
respectively, and $c$ is the light speed.
If the pre-supernova WR core have an angular velocity $\omega_C$ and
a radius R$_C$, then the angular momentum conservation leads to
$\omega_C$R$_C^2 \sim \omega_K$R$_K^2$. Because of the core co-rotation with
the orbit motion, $\omega_C \approx 2 \pi$/P$_{\rm orbit}$, then a Kerr BH
is produced if cP$_{\rm orbit}\lesssim$ 2$\pi$R$_C$(R$_C$ / R$_K$).
If R$_C \sim$ 0.5 R$_{\sun}$ and R$_K \sim$10$^7$ cm, then P$_{\rm orbit} 
\lesssim$ 7 hr, while the separation of the binary system, if the total mass 
$\sim$30M$_{\sun}$, is $\lesssim$6 R$_{\sun}$. Close WR+BH binaries show these 
properties. Such systems exist in nature, and some potential candidates are 
known, for example Cyg X-3 (WN3-7+ BH) with P$_{\rm orbit}$=0.2 days. These 
GRB progenitor systems might be observed as luminous, massive X-ray binaries. 

We can now estimate the FR of massive WR+BH binaries in the Milky Way.
Using the initial star-formation function for close binaries (Iben \& Tutukov
1984):
\begin{equation}
{\rm d^3N = 0.1\ \Bigl(\frac{da_{in}}{a_{in}}\Bigr)  
\ \Bigl(\frac{M_1}{M_{\odot}}\Bigr)^{-2.5}
d\Bigl(\frac{M_1}{\msun}\Bigr)\ dq \ \ {\rm yr}^{-1}},
\end{equation}
where a$_{\rm in}$ is the initial system semi-major axis, M$_1$ is the 
initial mass of the primary, and q=M$_2$/M$_1$. This function has been 
determined from observations in the solar neighborhood and implies a roughly
uniform distribution of a$_{\rm in}$ for 10$<$a$_{\rm in}$/R$_{\sun}<$10$^6$,
a mass distribution of the primaries close to the Salpeter IMF, and a uniform 
distribution of q.  If we adopt M$_1 \sim$ 25 M$_{\sun}$, 
da$_{\rm in}$/a$_{\rm in} \sim$ 0.5, and dq$\sim$0.3, the estimate of the Kerr 
BH formation rate in the Milky Way is $\sim$ 10$^{-4}$ yr$^{-1}$. 
This rate, whose uncertainty is about a factor 3, corresponds to a population 
of a few close binary WR progenitors of Kerr BHs at present in the Galaxy. 
The Kerr BH formation rate derived through the previous purely astronomical
arguments matches rather well with the formation rate we have inferred from GRBs. 
Thus, a self-consistent scenario supports the idea that {\it the GRB progenitors 
should be close binary massive WR stars, possibly with a massive BH companion}.
 
As well as the SF enshrouded in dense dust clouds, several other effects
can influence the shape difference between the GRB FRH and the 
cosmic SFRH obtained from rest-frame UV luminosity.  
Metallicity influences the star mass loss and evolution, and consequently
the energetic and collimation of GRB. 
Even the IMF and binary formation function may change with redshift. 
All these questions should be well understood before using the GRB FR as 
a tracer of the cosmic SFR.

\section{Conclusions}

In this work we have exploited the observational peak-flux distribution
for more than 3300 GRBs, and the redshift distribution for 220 GRBs
inferred from an empirical luminosity-indicator relationship in 
order to constrain {\it jointly} the LF and FRH of long GRBs. Our 
analysis allows us to draw the following conclusions: \\

-For single or double power law LFs, the case of non-evolving LF
fits poorly the data, while evolving LFs fit rather well both
the NPD and NZD. The best fits are obtained for an evolution where
luminosity scales as (1+z)$^{\delta}$, being $\delta={\rm 1.0} \pm {\rm 0.2}$.
This evolution increases the probability to observe GRBs from very high 
redshifts. The introduction of a dependence of \Liso\ on E$\rm_b$ (Eq. 4) 
has little effect on the models, the most important one being 
the improvement of the fit for the models with non-evolving LF.
The goodness of the fit of models with both the SPL and DPL evolving LFs 
are excellent. The quality of the fit for the former is slightly 
better than for the latter.

-The best models provide a GRB FRH that approximately resembles the 
observed cosmic SFRH, in particular if the potential contribution
of the SFR in the obscured regime is taken into account. The FRH rises
steeply from z=0 to z$\sim$2, and then increases gently 
toward higher redshifts. The results are qualitatively similar
when using a sample of 33 GRBs with known z and adequately
de-biased from selection effects. More data, in particular in the 
NZD, are necessary to explore in more detail the connection of GRB FR
to the cosmic SFR.

-For the SPL LF, the best slope from the fits is 
$\gamma = 1.57\pm 0.03$. In the understanding that the LF is
closely related to collimation effects, this result implies
an intermediate case between the universal  structured jet
model with  $\epsilon(\theta)\propto \theta^{-2}$ and the
quasi-universal Gaussian structured jet model. For the
uniform jet model, the jet angle distributions corresponding 
to the best model were calculated, giving an indicative range 
between 2$^\circ$ and 15$^\circ$ at z=1. 

-Our best models give a {\it true} (after collimation effect correction) 
GRB FR of $\sim$ 5 10$^{-5}$ yr$^{-1}$ in the Milky Way. Based on 
astronomical arguments we have argued that such a FR agrees with that 
of close binary systems consisting of a WR star and a possible
massive BH, with periods of hours. These systems are able to 
generate a massive Kerr BH after the 
SNIb/c explosion of the WR (helium) star. The observational counterparts of 
these potential GRB progenitors should be luminous X-ray binaries 
(e.g., Cyg X-3), which are estimated to be only a few at present 
in the Milky Way.

\acknowledgments

We thank J.S. Bloom, E. Ramirez-Ruiz and B.E. Stern for providing
us with data used in this paper,  and the anonymous referee
for helpful comments and questions. We also thank G. Malaspina for 
technical assistance. Support for this work was provided by CONACyT 
grant 33776-E to V.A. and by COFIN grant 2001022957/004 to G.G.

\clearpage
\begin{figure}
\vspace{18cm}
\includegraphics{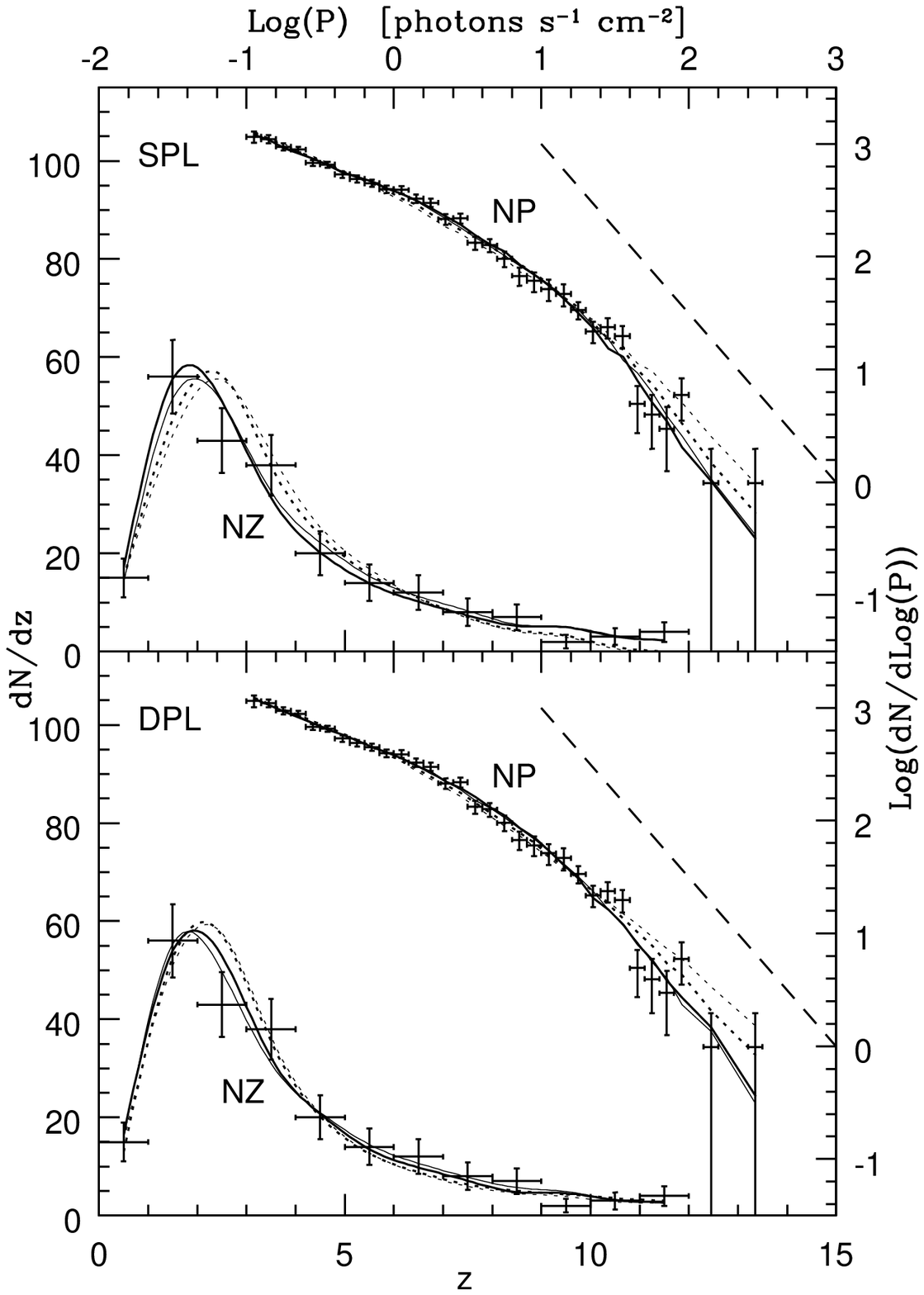}
\caption{{\it Top panel:} Peak flux differential distribution (NP) with 
the axis in the top-right part, and redshift differential distribution 
(NZ) with the axis in the bottom-left part, both for a SPL LF. Error 
bars show the NP data from SAH02 and the NZ data according to FR00, 
respectively. Dotted lines are for models without evolution (0), while 
solid lines are for models (E) with the evolution parameter $\delta$ 
optimized. Thin and thick lines identify the cases with E$\rm_b$=511 
keV (P) or E$\rm_b$ depending on \Liso according to Eq. (4) (case Q), 
respectively. Dashed straight line is the -3/2 uniform distribution 
(Euclidean) behavior in the NPD. 
{\it Bottom panel:} Same as in top panel but for a DPL LF.\label{fig1}}
\end{figure}

\clearpage

\begin{figure}
\vspace{18cm}
\includegraphics{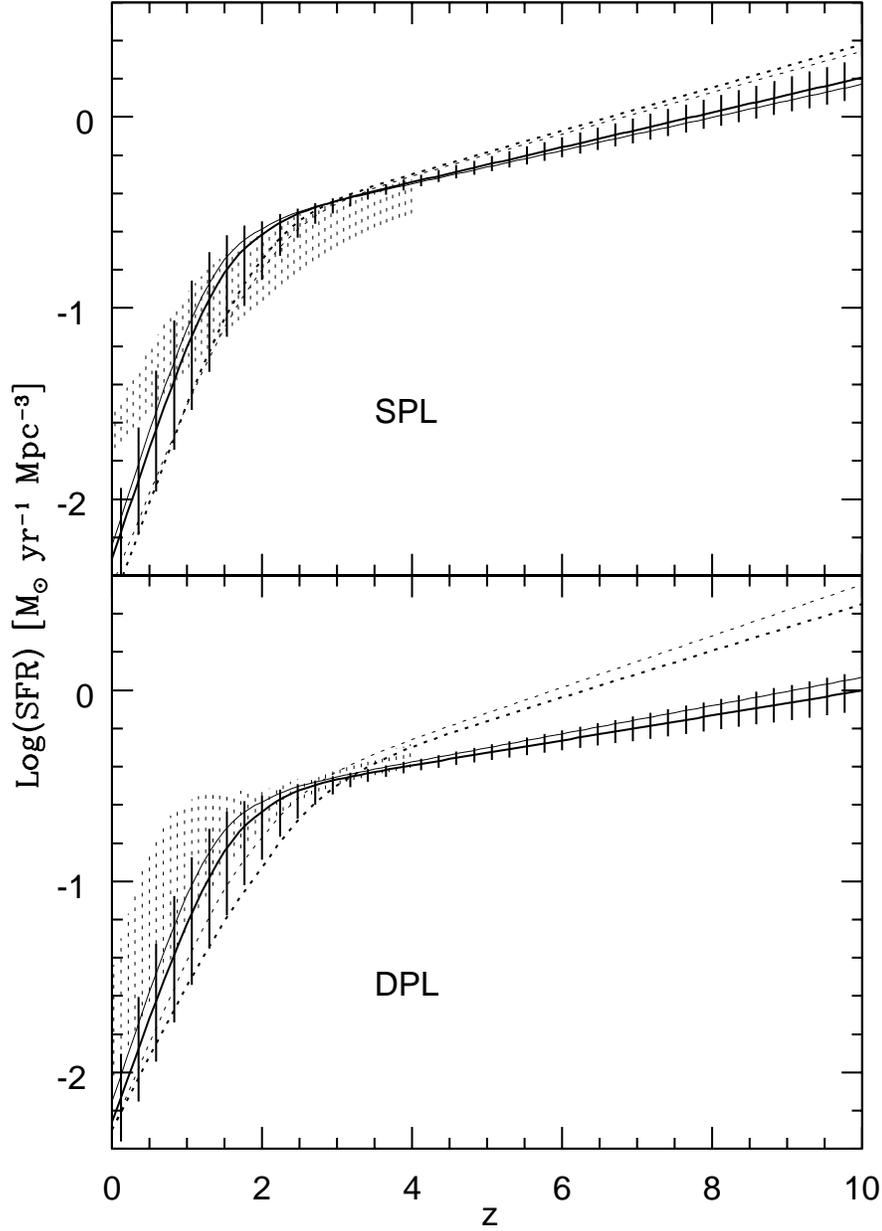}
\caption{SFRHs inferred from the GRB FRHs under an opportune normalization. 
Symbols and models are the same as in Fig. 1. Vertical segmented 
areas show the 1 $\sigma$ uncertainty for the evolutionary models. 
Shaded areas show the evolving models, including their uncertainties, 
for the NZD data corresponding to 33 GRBs with z known (PR03) and 
de-biased from selection effects. \label{fig2}}
\end{figure}

\clearpage

\begin{figure}

\plotone{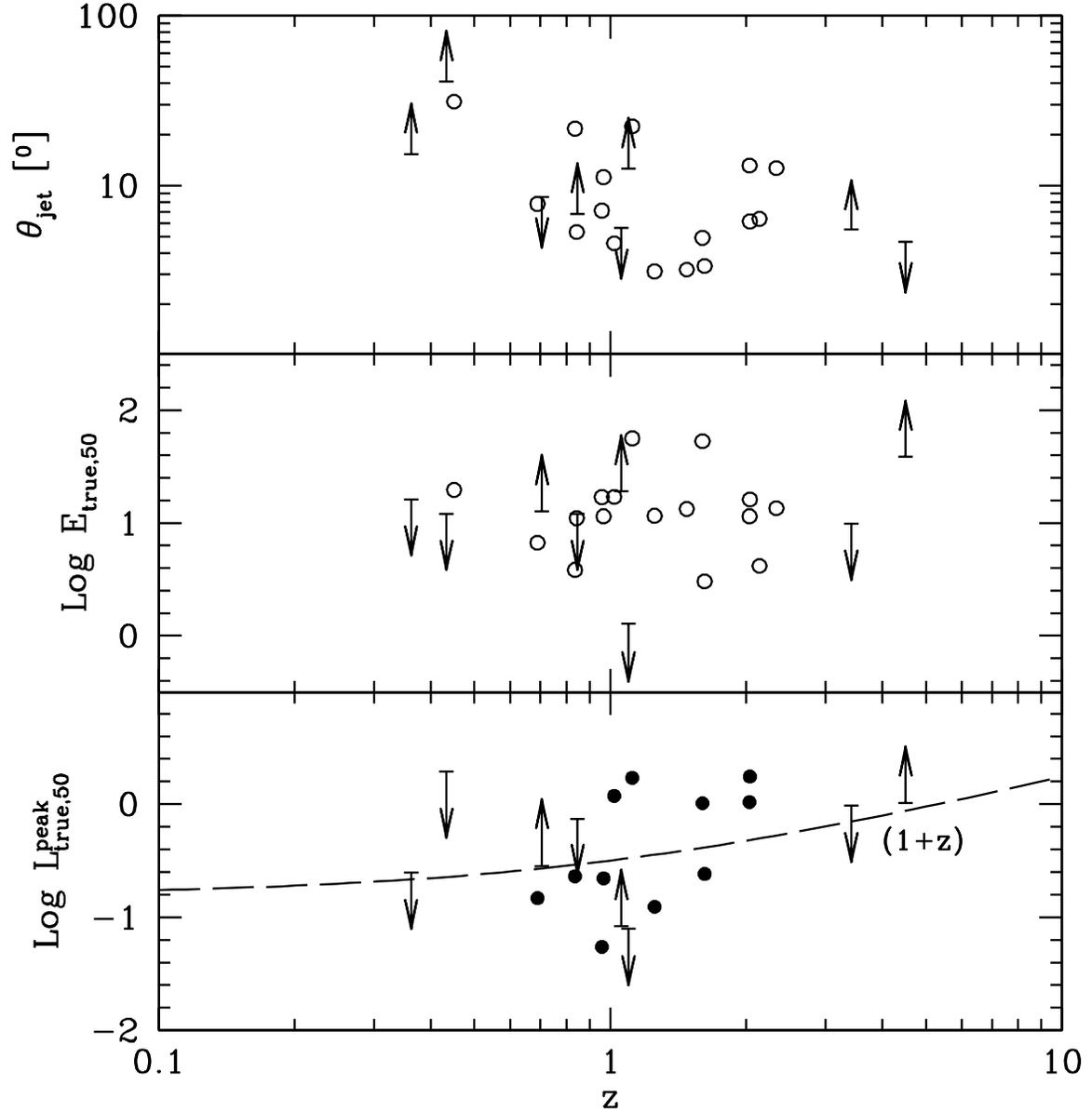}
\caption{From top to bottom: jet opening angle $\theta$, {\it true} emitted 
energy E$_{\rm true, 50}$ (in units of 10$^{50}$ erg) and {\it true} emitted 
peak luminosity L$^{\rm peak}_{\rm true, 50}$ (in units of $10^{50}$ \ergs), 
as a function of the redshift z.
The values (and limits, for a total of 24 data points) of the jet opening angles 
and energies are taken from Bloom et al. (2003), while the values 
from the true luminosities (19 points including limits) have been calculated using 
the peak count rate of the GRB listed in PR03. 
In the bottom panel the dashed lines is proportional to (1+z) to show that the mild 
evolution we find is compatible with existing data and the suggestion of a universal 
energy reservoir. \label{fig5}}

\end{figure}

\clearpage

\begin{figure}
\plotone{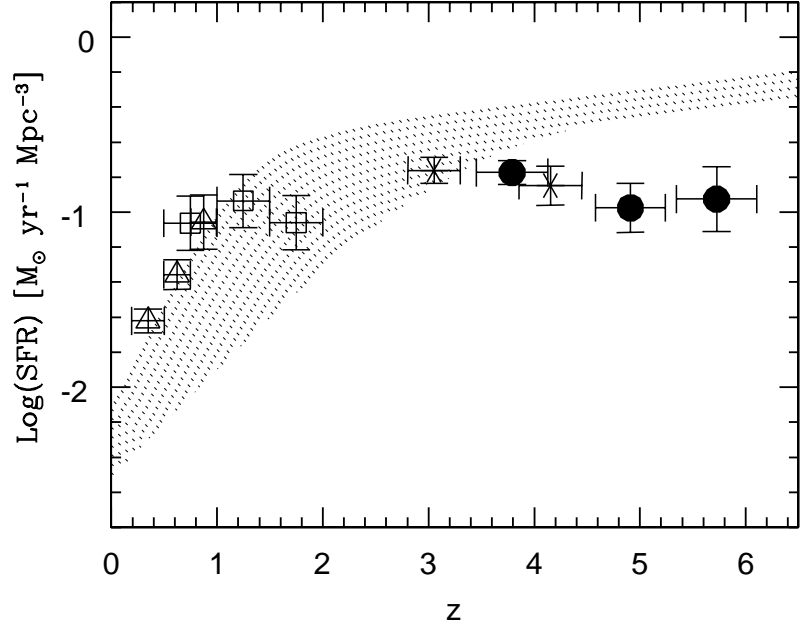}
\caption{Comparison between the observed cosmic SFRH traced by the rest-frame 
UV luminosity and corrected by dust obscuration from Giavalisco et al.
(2004) (dots with error bars) and the SFRH obtained from GRB FRH opportunely 
normalized (shaded area). The selected models include the uncertainties and 
correspond to $\delta$=1, and Eq. (4). 
\label{fig6}}
\end{figure}

\clearpage

\begin{table}
\begin{center}
\caption{\centerline{Best-fit parameters for the LF models}\label{tbl-1}} 
\begin{tabular}{ccccccccc}
\tableline\tableline

Model\tablenotemark{a} & 

$\gamma$ &

$\rm LogL_u$\tablenotemark{b} & 

~~~~~~~   &

Loga &

b &

c & 

$\rm \delta$ &

K\tablenotemark{c} \\ 

\tableline

SP0 & 1.58$\pm$0.04 & 4.06$\pm$0.07 & ~~~~~~~ & 1.9$\pm$0.2 & 2.3$\pm$0.4 & 0.25$\pm$0.04 & 

~~ & 10.4 \\

SPE & 1.53$\pm$0.03 &  3.3$\pm$0.1  & ~~~~~~~ & 1.7$\pm$0.3 & 2.7$\pm$0.6 & 0.20$\pm$0.03 & 

0.9$\pm$0.1 & 3.43 \\

\tableline

SQ0 & 1.59$\pm$0.03 & 3.96$\pm$0.05 & ~~~~~~~ & 2.0$\pm$0.2 & 2.5$\pm$0.4 & 0.26$\pm$0.03 & 

~~ & 11.6 \\

SQE & 1.57$\pm$0.03 &  3.3$\pm$0.1  & ~~~~~~~ & 1.8$\pm$0.2 & 2.8$\pm$0.6 & 0.21$\pm$0.03 & 

0.8$\pm$0.1 & 4.93 \\

\tableline

\end{tabular}

\begin{tabular}{ccccccccc}

Model\tablenotemark{a} &

$\gamma_1$ & 

$\rm LogL_b$\tablenotemark{b} & 

$\gamma_2$ &

Loga &

b &

c & 

$\delta$ &

K\tablenotemark{c} \\ 

\tableline

DP0 &  0.9$\pm$0.4  & 2.6$\pm$0.3 & 2.1$\pm$0.2 & 1.5$\pm$0.4 & 1.6$\pm$0.6 & 0.29$\pm$0.05 & 

~~ & 0.73 \\

DPE & 1.53$\pm$0.03 & 2.8$\pm$0.1 & 3.4$\pm$0.5 & 1.6$\pm$0.3 & 2.8$\pm$0.7 & 0.17$\pm$0.04 & 

1.2$\pm$0.1 & 2.85 \\

\tableline

DQ0 &  0.8$\pm$0.5  & 2.4$\pm$0.4 & 2.1$\pm$0.2 & 1.8$\pm$0.4 & 1.8$\pm$0.7 & 0.28$\pm$0.05 & 

~~ & 0.78 \\

DQE & 1.57$\pm$0.04 & 2.6$\pm$0.1 & 2.5$\pm$0.2 & 1.7$\pm$0.1 & 2.6$\pm$0.3 & 0.15$\pm$0.02 & 

1.1$\pm$0.1 & 4.83 \\

\tableline

\end{tabular}

\tablenotetext{a}{The symbol code means: S (SPL), D (DPL), P (E$\rm_b$=511\ keV), 
Q (E$\rm_b$ from Eq. 4), 0 (no evolution) and E (evolution)}  

\tablenotetext{b}{Luminosity unit 10$^{50}$ \ergs, energy range: 30-10000 keV}

\tablenotetext{c}{Number of GRBs per unit mass of gas transforming in stars
in units of 10$^{-8}$ M$_{\sun}^{-1}$}


\end{center}
\end{table}

\begin{table}
\begin{center}
\caption{\centerline{Statistical quality parameters of the models}\label{tbl-1}} 
\begin{tabular}{ccccccc}
\tableline\tableline

Model\tablenotemark{a} & 

$\chi^2$ &

Q &

$\chi_{\rm NP}^2$ &

$\chi_{\rm NZ}^2$ &
 
$\rm P_{\rm NP}^{\rm KS}$ &

$\rm P_{\rm NZ}^{\rm KS}$ \\ 

\tableline

SP0 & 84.3 & 2e-5 & 68.0 & 16.3 & 0.46 & 0.12 \\

SPE & 35.7 & 0.48 & 26.9 &  8.8 & 1e-6 & 1e-6 \\

\tableline

SQ0 & 56.6 & 0.02 & 42.8 & 13.8 & 3e-3 & 0.02 \\

SQE & 31.7 & 0.67 & 24.4 &  7.4 & 1e-6 & 3e-3 \\

\tableline







 



DP0 & 74.6 & 2e-4 & 64.3 & 10.3 & 0.26 & 0.02 \\

DPE & 35.5 & 0.44 & 28.0 & 7.5 & 1e-6 & 5e-5 \\

\tableline

DQ0 & 50.0 & 0.06 & 39.8 & 10.2 & 0.02 & 4e-3 \\

DQE & 36.6 & 0.40 & 28.3 & 8.3 & 1e-6 & 5e-5 \\

\tableline

\end{tabular}

\tablenotetext{a}{The same symbol code of table 1}

\end{center}
\end{table}

\end{document}